# Polarization Dynamics in Ferroelectrics: Insights Enabled by Machine Learning Molecular Dynamics

*Dongyu Bai[1#], Ri He[2#], Junxian Liu[1], and Liangzhi Kou[1]\**

1. School of Mechanical, Medical and Process Engineering, Queensland University of Technology, Brisbane, Queensland 4001, Australia
2. Key Laboratory of Magnetic Materials Devices & Zhejiang Province Key Laboratory of Magnetic Materials and Application Technology, Ningbo Institute of Materials Technology and Engineering, Chinese Academy of Sciences, Ningbo 315201, China

# Equal contribution

E-mail: *Liangzhi.kou@qut.edu.au*

# ABSTRACT

Ferroelectric materials with switchable spontaneous polarization underpin non-volatile memories, transistors, sensors, and emerging neuromorphic chips. Their performance and stability are governed by polarization dynamics and domain kinetics, making a microscopic understanding of these processes and precise atomic-level control of polarization domains key challenges for next-generation ferroelectric electronics. Due to the limitations of the characterization technology with atomic-level in experiment, high-precision atomic simulations become important. First-principles calculations are inherently limited in accessible length and time scales, making it difficult to capture the complex features of dynamic processes. Machine-learning molecular dynamics (MLMD) offers a compelling solution by encoding quantum-mechanical accuracy into force fields, thereby enabling large-scale dynamic simulations with near first-principles fidelity. This Perspective highlights the advantages of MLMD for simulating polarization switching, domain nucleation and migration, topological polar textures and curvature-driven ferroelectric phenomena, while providing a systematic overview of recent progress in these areas. We further discuss methodological challenges that limit predictive capability, including long-range electrostatics, coupled lattice-spin degrees of freedom in multiferroics, and data-efficient pre-training of large atomistic models. Corresponding advances in long-range-aware force fields, spin-dependent machine-learning models, and large-scale pre-training are expected to move MLMD toward a genuinely predictive framework for the design of ferroelectric and multiferroic materials.

# 1. Introduction

Ferroelectric materials are defined by a spontaneous electric polarization that can be reversibly switched under an external electric field, giving rise to a nonlinear polarization–electric field (P-E) hysteresis loop analogous to magnetic hysteresis. Since the first discovery of ferroelectricity in Rochelle salt in 1920 [1], the field has witnessed sustained progress in both fundamental insights and technological exploitation. Ferroelectric family also extends from classical bulk inorganic oxides [2-4], polymers [5-7], metal-organic framework (MOF) ferroelectrics [8-10], and thin films [11-12], to emerging two-dimensional (2D) layered systems [13-20]. Owing to their rich electromechanical properties such as piezoelectricity [21-23], flexoelectricity [24-25], and electro-optic effects [26-27], ferroelectrics underpin a broad spectrum of applications ranging from non-volatile memories [28-29], neuromorphic chip [30-33], photocatalysis [34-36], sensing [37], and electronic devices [38] (**Figure 1a**).

The multifunctionality of ferroelectrics originates from their bistable polarization states and the associated reversible polarization dynamics, which form the physical basis of device operation. For instance, binary logic states (0 and 1) in ferroelectric non-volatile memories are encoded by two stable polarization orientations [39], while polarization-controlled modulation of channel conductance enables ON and OFF switching in ferroelectric field-effect transistors [40]. Consequently, the performance of ferroelectric devices is intrinsically governed by polarization switching and its dynamics. These dynamic processes, involving of characteristic domain configurations and domain wall (DW) structures (**Figure 1b**), directly determine key device metrics such as switching speed, energy efficiency, and long-term reliability of devices. The data write/read operation speed, for example, is closely related to the nucleation barrier and DW mobility [41-42], as well as Schottky barriers [43-44], depolarization fields [45-46] and structural defects [47-48]. Reliability of the ferroelectric devices is dependent on defect-induced DW pinning and energy barriers of domain migration, which manifests macroscopically as polarization fatigue and progressive performance degradation [49-52]. In ferroelectric memristors or neuromorphic devices, input spikes (voltage pulses) induce domain wall motion and polarization continuous changes, thereby modulating the device's conductance—a direct analog to synaptic strength. This allows the device to perform both memory and processing functions simultaneously, akin to biological neurons [33, 53].

To study polarization dynamics and DW motion in ferroelectrics, advanced experimental techniques and theoretical modelling approaches have been developed and applied to address the critical issues, including high-resolution microscopy, first- and second-principles calculations [54-56], and empirical or semi-empirical simulation methods such as mesoscale phase-field modeling [57-58], molecular dynamics (MD), Monte Carlo simulations [59-65]. For example, phase-field simulations were used to reveal complex 180° DW configurations and interactions in ferroelectric Pb(Zr,Ti)O$_3$ (PZT) [66], second-principles calculations successfully reproduced experimentally observed domain nucleation mechanisms and growth kinetics in PZT [60], four-dimensional scanning transmission electron microscopy (4D-STEM) together with first-principles calculations uncovered in-plane topological vortex domain structures in twisted bilayer MoS$_2$ [67]. Despite these successes, capturing polarization dynamic processes still poses formidable challenges for either experiments or conventional simulations. The underlying reason is, polarization switching, DW nucleation, motion, and interaction are inherently dynamic processes occurring on ultrafast (picosecond–nanosecond) time scales and atomic-to-mesoscopic length scales, while conventional simulations and advanced experimental measurements are mainly for the static configurations at specific instance. For example, in aberration-corrected TEM, what is observed is not the signal of isolated individual atoms but the scattering intensity arising from atoms that overlap in projection along the beam direction [68-69]. The lack of depth information can introduce systematic errors in the quantitative determination of spatial parameters such as DW thickness and displacement [70-71]. In addition, high-resolution probes may introduce artifacts such as beam-induced charging, structural damage, or tip/beam written domains [72-75]. From a theoretical perspective, static first-principles calculations, while quantitatively accurate, are fundamentally limited by accessible system sizes and time scales, hindering their direct application to mesoscale DW networks and long-range collective phenomena. Although classical MD and phase-field models can bridge larger temporal and spatial regimes, their predictive power critically depends on the availability of high-fidelity interatomic potentials or accurate materials parameters. For newly emerging ferroelectric families, such as 2D ferroelectrics, sliding ferroelectrics [20, 76-78], and MOF-based systems, the construction and calibration of these models have become a major bottleneck. This process often requires extensive experimental input or iterative fitting to first-principles data within multiscale simulation-experiment feedback loops. Moreover, the continuum model, phase-field modelling cannot directly capture van der Waals (vdW) layered materials, particularly the atomic stacking registry and interlayer sliding associated with the vdW gaps.

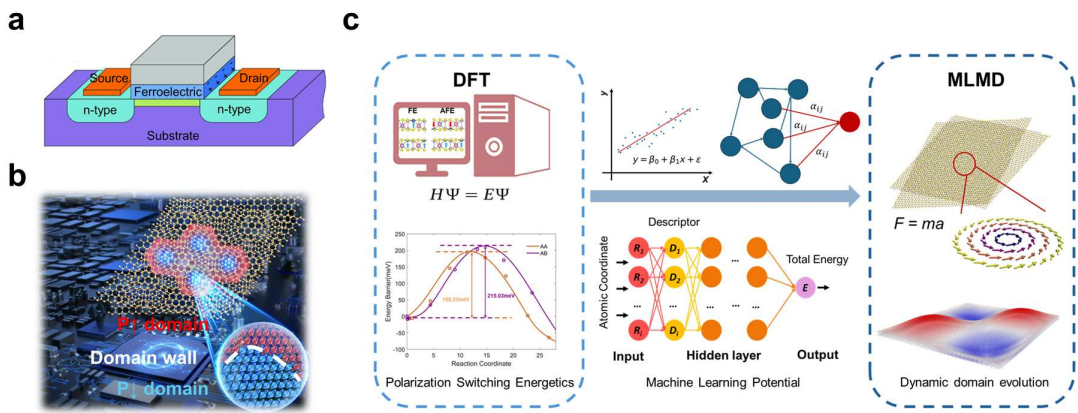

**Figure 1. From first-principles to MLMD simulations for ferroelectric studies.** (a) Schematic diagram of ferroelectric materials in Ferroelectric Field-Effect Transistor (FeFET) devices. (b) Domain and domain-wall configurations in twisted ferroelectric layers. (c) Roadmap from static DFT simulations to dynamic MLMD simulations, where MLFFs are trained based on DFT data as a bridge to enable the large-scale dynamic studies. The developed force fields are then employed in MLMD simulations to investigate polarization switching and domain evolution. The bubble structure shown in the MLMD section is adapted with permission from ref [25]. Copyright 2023, American Chemical Society.

In this context, MLMD has emerged as a powerful and feasible atomistic simulation paradigm for ferroelectric materials (**Figure 1c**). By training machine-learning force fields (MLFFs) on extensive density functional theory (DFT) datasets, MLMD delivers near first-principles accuracy for energies, forces, and stresses, while simultaneously scaling to system sizes of $10^3$–$10^5$ atoms and enabling long-time sampling under realistic temperature and electric-field conditions [79-81]. Moreover, MLFFs leverage data-driven representations of interatomic interactions, reducing reliance on hand-crafted empirical potentials and prescribed mechanism. Building on these features, MLMD can directly capture a range of key ferroelectric processes, including domain nucleation and growth [82-83], DW motion [84-85], defect DW interactions [77, 85], and polarization switching dynamics [86-87]. As a result, MLMD provides a direct bridge between atomistic mechanisms and mesoscale or device-level ferroelectric responses. Recent studies have further demonstrated its broad applicability across diverse ferroelectric materials, spanning conventional bulk oxides to emerging layered and 2D ferroelectrics [88-89].

With a particular focus on polarization dynamics and DW motion, this Perspective will first outline the fundamental MLMD frameworks and benchmark strategies relevant to ferroelectric systems. We then discuss how this approach enables the mapping of phase diagrams and the identification of polarization switching pathways, including energy landscapes, kinetic barriers, and their dependence on electric fields and mechanical strain. Recent MLMD advances in DW kinetics, polar-domain dynamics, and topological polarization textures will subsequently be reviewed, followed by an assessment of coupled polar–mechanical responses. We conclude by summarizing key challenges and outlining future strategies toward predictive, transferable, and experimentally guided MLMD for ferroelectric materials.

## 2. MLMD: From Concept to Implementation

To understand why MLMD is particularly well suited for studying polarization dynamics, it is instructive to first revisit the microscopic origin of polarization. Polarization in ferroelectric materials arises from the relative displacement between the centres of positive and negative charges within the crystal lattice. As a result, the formation of polarization domains and the evolution of DWs are fundamentally governed by ionic motions and local structural distortions. Capturing polarization dynamics therefore requires direct, time-resolved access to atomic-scale trajectories during domain nucleation, propagation, interaction, and switching. MD simulations provide a natural and powerful framework for describing such kinetic processes by explicitly resolving atomic motions in large-scale systems over extended time scales. By integrating with machine-learning techniques, MLMD can faithfully describe complex polar energetics and forces, while simultaneously accessing length and time scales that are inaccessible to direct first-principles simulations. MLMD therefore offers unprecedented opportunities to interrogate polarization domain dynamics, DW kinetics, and field-driven switching processes under realistic thermodynamic and electromechanical conditions. As such, it is emerging as a critical tool for advancing the fundamental understanding of ferroelectric behaviours and for guiding the rational design of ferroelectric materials and devices. In this section, we introduce the core principles of MLMD and outline how the methodology is implemented to address polarization dynamics in ferroelectric systems.

### 2.1. Basic Concept of MLMD

In MD simulations, all atoms are treated as rigid bodies, and their trajectories are governed by Newton's second law: $F_i = m_i \frac{d^2 r_i}{dt^2}$. Since electronic structure information is not included, the force $F_i$ acting on atom $i$ cannot be directly obtained as in DFT simulations. Instead, it is derived from an 3N-dimensional function of atomic positions in the configuration space $F_i = -\partial U/\partial r_i$, where $U = \sum_i U_i = U(r_{i1}, r_{i2}, ..., r_{in})$ is commonly referred to as the potential energy surface (PES). Therefore, the accuracy of MD simulations is entirely determined by the PES. In classic MD simulations, the PES is obtained by parameterizing the atomic environment such as bond lengths, bond angles, and reaction parameters [90-93] and embedding it into appropriate physical equations. The associated parameters are then fitted against small sets of DFT calculations or experimental data to obtain the corresponding potentials or force fields. Such force field highly depends on the physical intuition and prior experiences, rendering them generally non-universal with unguaranteed accuracy and reliability.

To address the intrinsic limitations of traditional potential, machine learning force field (MLFF) is trained on data-driven models to learn the mapping from atomic configurations to energies (and forces). To enable a potential trained on configurations containing only tens of atoms to be transferable to systems with millions of atoms, MLFF trainings typically adopt the following approximation: the total energy of the system is expressed as a sum of per-atom local energy contributions, where each contribution depends only on the local environment within a cut-off radius. To improve the accuracy and transferability of the potential energy surface while enhancing computational efficiency, the local environment representation is usually required to be invariant to translation, rotation, and permutation [94-97]. MLFFs are typically trained on datasets labelled by high-accuracy first-principles calculations, with the training configurations systematically selected to sample the regions of configuration space relevant to the target processes, see **Figure 2a**. Within this framework, the energy mapping is learned by flexible function approximators. Consequently, the predictive performance and transferability of MLFFs are inherently data-driven and rigorously established through in-domain validation, complemented by boundary tests designed to identify overfitting and unreliable extrapolation [98].

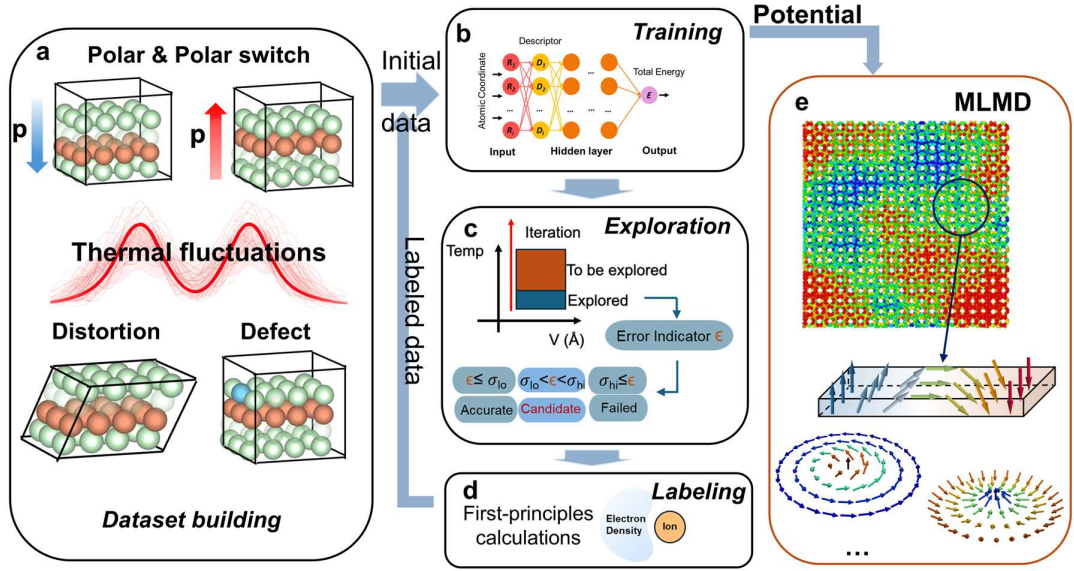

**Figure 2. Active learning workflow for ML potential development (Represented by DP-GEN [99]).** (a) Initial dataset building, representative configurations including different polarization states, thermal fluctuations, local distortions, and defects need to be covered. (b) Schematic diagram of neutral network for MLFFs training. (c) MLMD exploration under predefined conditions, and the model deviation of each sampled configuration is evaluated from the prediction disagreement among the MLFFs. Configurations with intermediate deviations are selected for labelling, while those with small deviations are considered reliable and those with excessively large deviations are discarded. (d) Configurations labelling by first-principles calculations for the next iteration. (e) large-scale MLMD simulations based on the trained MLFFs for complex polar-domain dynamics and thermodynamics studies.

## 2.2. MLFF Developments in Ferroelectrics

Early ideas for applying machine learning to large-scale molecular simulation date back to the 1990s, but the field was reshaped in 2007 by Behler and Parrinello [94], who first combined atomic environment descriptors with neural networks to approximate the PES. The structural representation and computational efficiency of MLFFs are significantly improved with the introduced descriptors (SOAP [95], Zernike [100], ACE [101], and SNAP [97]) and advanced architectures (DNNs [79, 102-103], GNNs [104-105], CNNs [106-107]) in recent years. These methodological innovations have enabled near first-principles molecular simulations at greatly reduced computational cost, making MLFFs increasingly practical for complex ferroelectric materials. In this section, taking DP-GEN as an example, we outline a typical active-learning workflow for constructing MLFFs for the constructing and deployment in ferroelectric research, and summarize the key steps for building PES models that are accurate and scalable across extended length and time scales.

The first step for MLFF training is to build training dataset. Polarization-related processes (e.g., polarization switching, domain nucleation, and DW motion) need to be identified, thereby defining the relevant regions of configuration space that require adequate coverage. Representative structural snapshots within these key regions are then generated from DFT calculations to establish the foundational dataset. For ferroelectrics, the dataset should include the asymmetric polar ground states and their characteristic local distortions; depending on the specific dynamic phenomena of interest, it will be further augmented with nonequilibrium configurations relevant to the target dynamics. Typical examples include intermediate states along representative switching pathways [83, 86, 108]; strain/ stress-induced distorted structures [109-111], thermally sampled structures at multiple temperatures, and local environments associated with defects or interfaces [77, 85, 89]. Following these principles, the initial dataset can be obtained via targeted structural perturbations, short ab initio molecular dynamics (AIMD) sampling at varied temperatures or nudged elastic band (NEB) calculations along candidate switching coordinates, ensuring physically meaningful coverage of the configuration space (**Figure 2a**).

The initial dataset is often static and may not adequately sample the vast configurational space of complex systems, leading to poor transferability. Manual construction of these configurations is also a tedious task, particularly for the states far from the bistable polar states. The concurrent learning, also known as active learning or on-the-fly learning, has emerged as a powerful paradigm to overcome this limitation. As a classic active learning workflow in the MLFF field, DP-GEN [99] follows an iterative "Training–Exploration–Labeling" loop (**Figure 2**). An ensemble of MLFFs trained on an initial dataset is first deployed to explore configurations under diverse thermodynamic and driving conditions. The sampled structures are then screened using a model-deviation criterion: configurations with intermediate deviation are selected for additional first-principles labelling and incorporated into the dataset, whereas excessively deviating samples are often treated as unphysical configurations and discarded. The models are subsequently retrained, and the cycle repeats until uncertainties and errors are reduced to an acceptable level across the targeted region of configuration space. Beyond DP-GEN, commonly used active-learning strategies also include RSS-driven incremental fitting workflows such as AutoPLEX [112], Bayesian uncertainty-driven on-the-fly active learning [113-114], and D-optimality selection schemes for linear-parameter potentials [115]. Those approaches enable the automated, iterative, and targeted construction of an optimal training dataset.

Based on the generated dataset, MLFF will be trained using mature and user-friendly software packages developed in recent years. Representative examples include deep neural network potentials such as Gaussian approximation potentials (GAP) [95], Behler–Parrinello neural networks (BPNN) [94], and DeepMD-kit [79, 116]; and graph neural network potentials that explicitly encode geometric symmetries (e.g., SchNetPack, NequIP and MACE [80, 104, 106, 117-118]); and GPU-optimized molecular dynamics engines with built-in MLFFs (e.g., NEP in GPUMD [119]).

The reliability and accuracy of the trained potential will be validated against benchmarks closely related to polarization dynamics, including the relative stability of polar states, elastic/phonon properties, switching barriers along representative pathways, DW structures and energies, and field/strain responses. Agreement with these benchmarks provides quantitative confidence in applying the MLFF to the investigation of DW kinetics and polarization switching mechanisms.

### 2.3. Implementing MLMD for Polarization Dynamics

With the trained MLFFs, MLMD simulations now can be conducted to probe polarization dynamics over extended spatial and temporal scales, by explicitly monitoring the atomic trajectory within in a framework analogous to traditional MD simulations. Notably, polarization is not typically a standard direct output of the potential model itself. Instead, polarization-related information is obtained through two main routes. The first is to learn polarization explicitly within the ML framework, by treating dipole moment as supervised targets and training an independent property network in parallel with the energy/force model [120-121]. The second relies on a post-processing of MLMD trajectories, where polarization is reconstructed from atomic displacements ($u_i$) or configurations using Born effective charges ($z_i^*$) obtained from first-principles calculations, according to $\vec{p} = \frac{1}{V}\sum_i z_i^* u_i$, with $V$ denoting the cell volume. The reconstructed polarization can be benchmarked against a limited set of Berry-phase polarization results [83, 85, 89, 110-111]. In practice, the post-processing route is more widely adopted because generating robust polarization labels for direct supervised training is computationally expensive, and polarization is highly sensitive to long-range electrostatics and boundary-condition effects.

## 3. Ferroelectric Research Progress enabled by MLMD

In this section, we review key developments and representative studies enabled by MLMD simulations from four interconnected perspectives: (i) polarization switching dynamics, (ii) polar DW kinetics, (iii) topological polar structures and (iv) polar-mechanical coupling.

Together, these advances highlight the transformative role of MLMD in reshaping how ferroelectric dynamics are interrogated, quantified, and ultimately controlled.

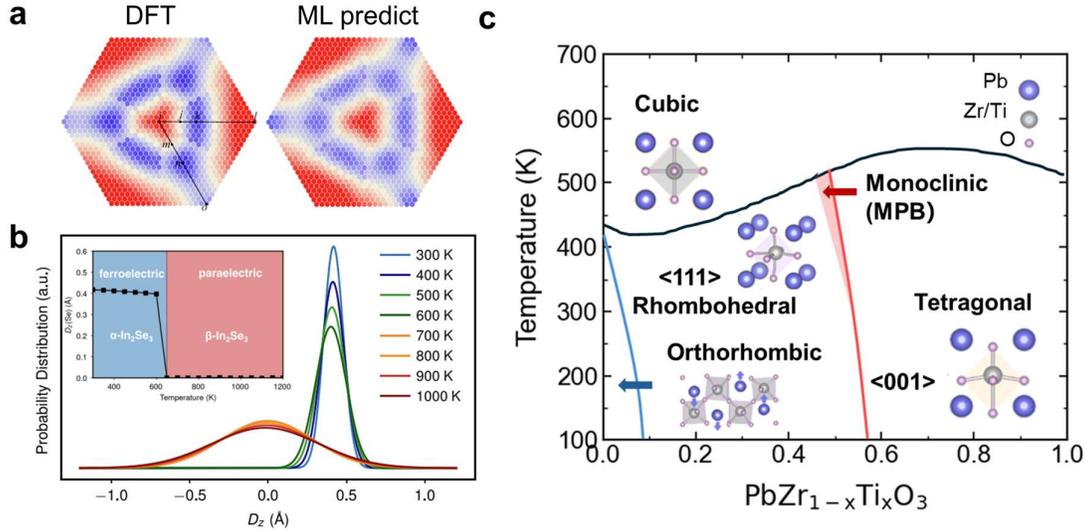

**Figure 3. MLMD predictions of ferroelectric switching and phase behavior.** (a) Comparison of the PES for in-plane sliding of the central Se sublayer in monolayer $In_2Se_3$ from the MLMD and DFT, with the minimum-energy switching pathway indicated by the solid line. (b) Temperature-dependent probability distributions of the central Se-sublayer displacement, Dz, from MLMD simulations. Reprinted by permission [86]. Copyright 2021, American Physical Society. (c) Temperature-composition phase diagram of PZT from MLMD predictions. Adapted from Refs [88, 122] with permission. Copyright 2024, American Physical Society.

## 3.1 Polarization Switching Dynamics

The defining signature of ferroelectricity is reversible polarization switching, which is governed by the microscopic mechanisms of ionic off-centering [13-14, 83, 123], and molecular rotations [10], interlayer sliding [20, 76-78]. These mechanisms dictate the kinetic pathways of switching and are intrinsically linked to structural distortions, as seen in $d^0$ $ABO_3$ ferroelectric perovskite where polarization reorientation is driven by local rhombohedral–tetragonal transitions [124]. Consequently, a coherent framework that bridges structural evolution with polarization dynamics is central to modern ferroelectric science. MLMD has emerged as a potent instrument for elucidating these structural dynamics [87-88], due to its high fidelity to track switching kinetics and phase evolution in large-scale simulations. In ferroelectric $In_2Se_3$ monolayer, the switching pathway and the energetic landscape among different phases are accurately reproduced (**Figure 3a**) with developed MLFFs, which ensures the accuracy and reliability of large-scale dynamic simulations. Moving beyond static barriers, structural evolution is captured with MLMD simulations under realistic conditions. As temperature rises, the probability distribution of the out-of-plane displacement ($D_z$) narrows toward zero at ~650 K, signalling a transition from the ferroelectric α-phase to the paraelectric β-phase [86] (**Figure 3b**). In single-component $SrTiO_3$, MLMD accurately captures the cooling-driven structural transition driven by antiferrodistortive (AFD) anti-phase rotations of oxygen octahedra. By mapping the spatial distribution of the AFD order parameter, simulations trace the establishment of long-range order as the system cools from 200 K to 50 K, enabling the construction of a refined in-plane biaxial strain–temperature phase diagram [125]. For complex solid perovskites $PbZr_{1-x}Ti_xO_3$ (PZT), MLMD provides critical corrections to the established phase diagram (**Figure 3c**). It is found that PZT retains $R_{3c}$ symmetry across the rhombohedral region, transitioning directly to cubic without an intermediate $R_{3m}$ phase due to persistent octahedral tilts. The polydomain nanostructures are formed near the morphotropic phase boundary (MPB), weak electric fields can easily induce polarization rotation [88]. The phase diagram can be markedly reshaped with the degree of cation ordering, providing new theoretical design principles for optimizing the electromechanical performance of PZT [122].

Beyond its capacity to track dynamic structural evolution, large-scale MLMD can also reveal emerging mechanisms and fundamental characteristics of ferroelectric materials through finite-temperature dynamical simulations, which are often obscured in smaller-scale static DFT calculations. Taking $CuInP_2S_6$ (CIPS) as an example, its out-of-plane polarization mainly originates from the off-center displacement of Cu ions. First-principles calculations and experimental measurements indicate that Cu displacement leads to a quadruple-well energy landscape with local energy minima and complex switching dynamics [126-127]. With the developed MLFFs, the energy profile together with fundamental properties of CIPS (e.g., lattice constants and phonon spectra) can be well reproduced with deep potential model, indicating the accuracy and reliability of the method. The mechanism and energy barrier of out-of-plane polarization switching are then accurately revealed from large-scale MLMD, where Cu dipoles hop between the upper and lower sites. Moreover, a ferroelectric–paraelectric transition is predicted at temperature of Tc≈340 K, which is in good agreement with the experimentally reported value of ~315 K [83, 128]. Monolayer bismuth (Bi) presents a distinct challenge, exhibiting a far greater number of metastable configurations compared to other Group-V monolayers (P, As, and Sb), resulting in a highly convoluted PES [129]. The material's unique electronic structure, characterized by strong orbital hybridization and significant long-range interactions, makes obtaining reliable large-scale dynamical descriptions particularly difficult. To address this, force fields based on Message-Passing Neural Networks (MPNN) [130] have been employed to capture the subtle, non-local variations in the atomic environment. Subsequent simulations reveal that ferroelectric polarization in monolayer Bi emerges upon cooling through a cooperative mechanism involving spontaneous out-of-plane displacements and in-plane shear distortions [129].

In summary, MLMD simulations bridge a critical gap in ferroelectric research. They not only reproduce static fundamentals (energy barriers, energetic landscapes) with first-principles accuracy but also capture dynamic structural evolution in supercells containing thousands of atoms under realistic temperature and pressure conditions. This capability effectively mitigates finite-size artifacts and reduces computational costs by orders of magnitude compared to traditional *ab initio* methods.

### 3.2. Domain Wall Evolution and Kinetics

In the majority of ferroelectric materials, polarization switching does not occur via the coherent reversal of an entire domain. Instead, it proceeds through a more energetically favourable pathway involving domain nucleation followed by DW migration, a process that significantly lowers the kinetic barrier compared to uniform reversal [59, 131-133]. The critical role of DWs is particularly evident in α-$In_2Se_3$ monolayer. In a single-domain configuration, an applied in-plane electric field fails to induce reversible ferroelectric switching, triggering instead an irreversible α→β′ structural transition. However, the reversibility can be recovered in the presence of pre-existing DWs. This recovery is attributed to the local breaking of threefold symmetry of $In_2Se_3$ lattice at the DW interface, which enables the in-plane field to directly drive ferroelectric switching. In these configurations, DW motion exhibits stochastic, avalanche-like dynamics that conform to a characteristic creep law (**Figures 4a,b**) [134]. The mechanism of polarization reversal evolves significantly in multilayer (2H) α-$In_2Se_3$ where switching proceeds in a discrete, layer-by-layer fashion. Under an applied electric field, an in-plane DW characterized by a nonpolar and β-like intermediate layer develops and propagates sequentially from one layer to the next under an applied electric field. This buffer layer effectively reduces the large activation barrier associated with direct junctions between oppositely polarized layers. In complex multidomain configurations, the kinetics are further governed by the hierarchy of DW motion: out-of-plane DWs migrate first, subsequently guiding the advancement of in-plane DWs. Because these distinct DW types possess disparate migration barriers, the overall switching kinetics are highly anisotropic, offering a degree of freedom for the precise control of domain architecture [135].

In-plane anisotropy also manifests through DWs oriented along different crystallographic directions. In monolayer Bi, 180° DWs exhibit significantly higher mobility than 90° DWs [129, 136]. This disparity arises from their distinct polarization reversal mechanisms: while 90° DW migration involves significant lattice distortion, 180° DW motion proceeds via a synchronous buckling and shear mechanism that preserves the integrity of chemical bonds, resulting in a much lower kinetic barrier. The competition between these kinetic modes drives the emergence of complex architectures. The intertwined multidomain structures in monolayer Bi typically consist of domain pairs connected by highly mobile, charged 180° DWs and separated by slower, conventional 90° DWs. This differential mobility leads to the formation of an interstitial, "checkerboard"-like multidomain pattern. At the junction points where multiple domain walls intersect, the continuous rotation of the polarization vector spontaneously generates a series of topologically protected polar vortices [136] (**Figure 4c**). It is important to note that most current MLMD studies on these systems have not explicitly included long-range electrostatic corrections. However, DFT benchmarks indicate that electrostatics contribute substantially (approximately 10–15% in the case of monolayer Bi) to the total DW energy. Since these interactions fundamentally govern the stability of charged DWs and the core structure of vortices, incorporating long-range-aware ML architectures represents a critical frontier for future methodological development to ensure quantitative predict ability.

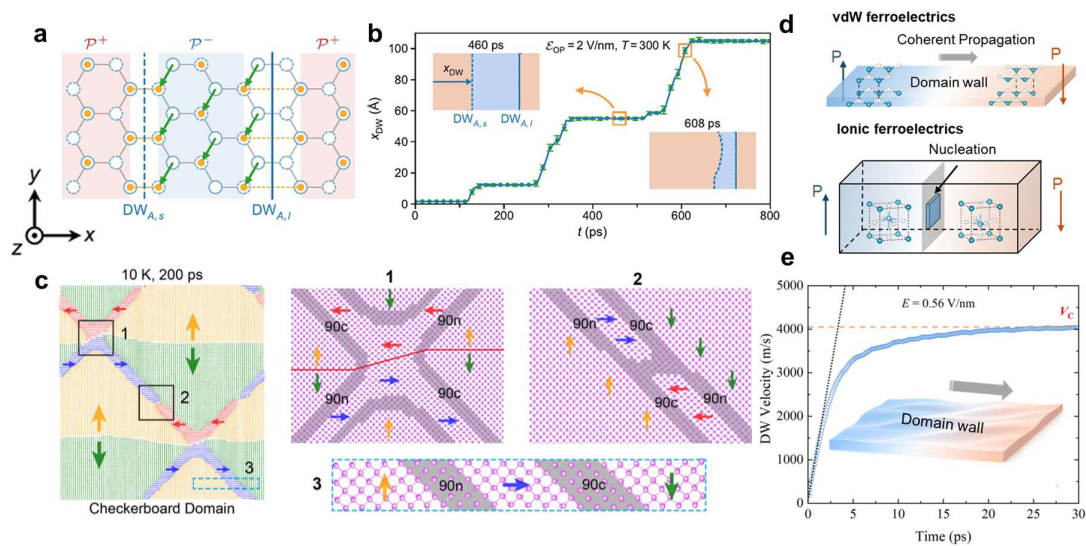

**Figure 4. Domain-wall structures, migration mechanisms, and dynamic responses in ferroelectrics from MLMD simulations.** (a) Two types of one-dimensional DWs ($DW_{A,s}$ & $DW_{A,f}$), induced by the central Se atoms displacement of monolayer α-$In_2Se_3$, which break the $C_3$ symmetry of single-domain structure. (b) Time evolution of $DW_{A,s}$ under internal electric field. Reprinted by permission [134]. Copyright 2025. American Chemical Society. (c) Atomistic structure of a checkerboard domain in a monolayer Bi structure.

Arrows denote polarization directions. Three boxed areas (1–3) mark representative regions. Gray areas and red solid lines indicate 90° domain walls and charged 180° domain walls, respectively. Reprinted by permission [136]. Copyright 2025. American Physical Society. **(d)** Comparison of DW transport mechanisms: in sliding ferroelectrics, DW motion is mediated by relatively weak interlayer vdW coupling, whereas in conventional ferroelectrics it is transmitted through strong ionic bond. **(e)** DW motion in sliding 3R-MoS$_2$ under an external field (0.56 V/nm), the propagation velocity approaches to the in-plane transverse acoustic phonon velocity $v_c$ of the material. Adapted from Refs [84] with permission. Copyright 2025, American Physical Society.

The presence of multiple polymorphs further complicates domain kinetics, underscoring the advantages of MLMD in capturing these complex dynamical processes. Taking monolayer Ga$_2$O$_3$ as an example, *Zhao et al.* employed a Gaussian Approximation Potential (GAP) [95] trained on diverse 2D configurations to bridge the gap between static DFT pathways and large-scale dynamics. Their MLMD simulations uncovered two fundamentally different nucleation-and-growth kinetics; Single-Atom Hopping: In the FE-ZB′ phase, switching is initiated by the thermally activated hopping of individual central-layer oxygen atoms. This leads to the nucleation of narrow domains that expand outward. Concerted Line-wise Migration: In contrast, the transition from FE-ZB′ to ML-β is dominated by the simultaneous movement of entire rows of oxygen atoms. Isolated displacements in this phase are energetically unstable, forcing the system into a more collective, "concerted" switching mode [82]. Similar approaches are also applied to antiferroelectric systems to probe domain structures and switching dynamics, even the DWs respond to external fields in ways that differ markedly from their ferroelectric counterparts. In PbZrO$_3$ (PZO) with multidomain configurations separated by 90° antiferroelectric DWs, externally applied fields can drive antiferroelectric to ferroelectric (AFE → FE) transition where successive dipole reorientates within the domains, but DWs remain largely immobile. This is in stark contrast to the DW propagation dominated switching typically observed in ferroelectric multidomain systems [137].

Unlike traditional ferroelectrics where polarization stems from intra-layer ionic off-centering, sliding ferroelectricity arises from symmetry breaking induced by the relative lateral displacement between van der Waals layers [20, 76-78] (**Figure 4d**). In these systems, the DWs are essentially boundaries between different stacking orders (e.g., AB vs. BA). Because the polarization $P$ depends directly on the interlayer relative ionic displacement $u$, the time-dependent evolution of $u$ can be accurately described by MLMD simulations, which is thus able to offer powerful insight into DW mediated switching in these complex 2D systems. A representative example is bilayer 3R-MoS$_2$, in which experiments report fatigue-free and ultrafast polarization switching: the switchable polarization shows negligible degradation over more than $10^6$ cycles, and the DW propagates in a soliton-like manner with velocities of ∼$10^3$ m/s [77]. Corresponding MLMD simulations based on defective bilayer $MoS_2$ models reveal the mechanism of fatigue-resistant behaviour as immobile charged S defects in single $MoS_2$ layer and defect-induced pinning of DW motion. The exceptionally low sliding switching barrier enables stable cycling over prolonged operation. More interestingly, it is found that the DW can be continuously accelerated under a sustained external field and exhibits "relativistic-like" dynamics, its velocity saturates as it approaches the characteristic in-plane transverse acoustic phonon velocity of ∼ $4×10^3$ m/s [84] (**Figure 4e**). Ultrafast DW motion in sliding ferroelectrics has also been experimentally confirmed in bilayer h-BN, with reported velocities reaching ∼ 1,000 m/s [138]. This behaviour is consistent with the fact that DW motion in sliding ferroelectrics is mediated primarily by interlayer shear and transmitted through relatively weak vdW coupling, which can substantially reduce energy dissipation during DW propagation compared with conventional ferroelectrics, where DW motion is transmitted by an ionic bond [139]. Consequently, ultrafast switching in sliding ferroelectrics such as 3R-$MoS_2$ and h-BN [85] does not primarily follow the conventional picture of thermally activated barrier crossing; instead, it is better described by the coherent propagation of wavelike DWs. In this regime, thermally induced ripples/corrugation at elevated temperatures increases the effective friction against DW sliding, such that cooling can enhance the mobility of these superlubric DWs [140]. Ultrafast, low latency, and low energy polarization switching in sliding ferroelectrics is expected to improve key device metrics such as write/read speed and cycling endurance, thereby supporting the development of scalable, high-density ultrafast ferroelectric memory devices.

### 3.3. Topological Polar Structures

Between two neighbouring domains, DWs will bend and evolve into flux-closure patterns similar to those observed in magnetic materials, when being constrained by finite sizes, boundary conditions, and external fields [141]. DW networks can further evolve into stable, complex polar topological textures such as polarization vortices and polar skyrmions [142-144] under the competing influences of depolarization, elastic interactions, and interfacial/interlayer couplings. These topological polar structures will be nucleated, moved, and annihilated by applied electric fields, offering a path to non-volatile memory with significantly lower energy consumption compared to magnetic skyrmion-based or conventional devices. Due to its importance of device applications, a diverse array of topological polar structures has been reported in nanoscale ferroelectrics, ranging from long-range ordered vortex–antivortex arrays in $(PbTiO_3)_n/(SrTiO_3)_n$ superlattice [142], to 2D ferroelectric vortex patterns in twisted freestanding $BaTiO_3$ layers [143], and stable in-plane chiral vortex domains in twisted bilayer $MoS_2$ [67] (see **Figure 5a**), and so on.

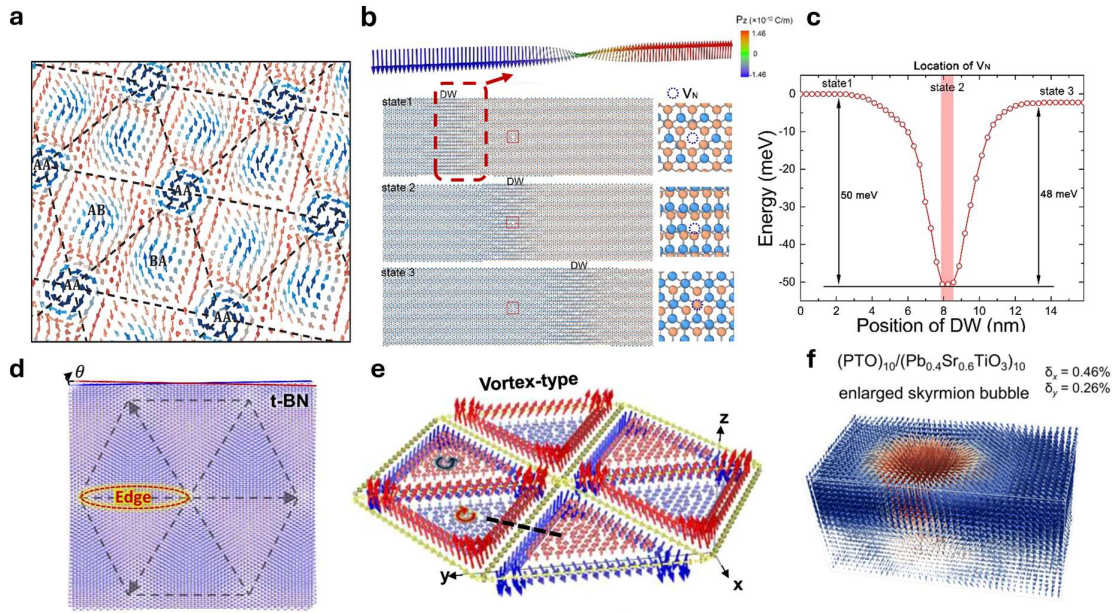

**Figure 5. Complex polar topological textures in ferroelectrics revealed by MLMD.** (a) Polarization distribution in $MoS_2$ moiré superlattice. Reprinted by permission [67]. Copyright 2024. The American Association for the Advancement of Science. (b) Atomic configurations of the bilayer h-BN system as the 0° DW approaches, interacts with, and moves away from the $V_N$ at different positions. Adapted by permission [85]. (c) Energy profile of the 0° DW as a function of its distance from the $V_N$. The energy minimum near the vacancy indicates a pinning effect induced by the defect. Reprinted by permission [85]. Copyright 2023. Acta Materialia Inc. Published by Elsevier Ltd. (d) Top view of relaxed atomic structure of twisted h-BN (t-BN). (e) Perspective view of the moiré structure, clockwise and counterclockwise meron and antimeron textures are observed, out-of-plane polarization at the edges is indicated by blue and red colors. Reprinted by licensed under CC BY 4.0 [145]. Copyright 2025, Springer Nature. (f) Skyrmion-bubble structures in ferroelectric (PbTiO3)10/(SrTiO3)10 superlattice by Pb doping in the $SrTiO_3$ layers. Reprinted by permission [146]. Copyright 2024, American Physical Society.

Distinct from static DFT calculations or classical MD simulations, MLMD moves beyond simple DW motion tracking and enables identification of the underlying "design rules" governing complex polar topologies in sliding- and twist-induced ferroelectrics. In twisted vdW bilayers, moiré superlattices give rise to spatially modulated polarization textures [19, 147]. However, the stability of these topological polar structures has been questioned based on static predictions in the absence of an external electric field [148-149] owing to small switching barriers and weak interlayer vdW interactions. This apparent contradiction is partially resolved by MLMD dynamic simulations. In ideal, defect-free h-BN bilayer, *He et al.* found that the macroscopic polarization rapidly decays to zero once the external electric field is removed, confirming the intrinsically ultrafast DW dynamics. Importantly, however, introducing even a small density of structural defects markedly increases the DW migration barrier. Strong DW pinning suppresses domain-wall motion, thereby significantly stabilizing the topological polar structures (**Figure 5b,c**). These results provide a rational explanation for how ferroelectric hysteresis can emerge in moiré superlattices despite their intrinsically low energy barriers and rapid switching kinetics [85]. Nevertheless, defect-induced DW pinning is not the sole stabilization mechanism in twisted bilayers. In an alternative study by *Fan et al.*, a different intrinsic mechanism was proposed for twisted bilayer h-BN [145]. The stable polarization was attributed primarily to coupling between sliding ferroelectricity and the piezoelectric response. At the boundaries of AB/BA triangular moiré domains, this coupling induces and stabilizes DW structures with complex edge-polarization topology (**Figure 5d,e**). By combining experimental characterization with MLMD simulations, the authors demonstrated that the edges of AB/BA triangular DWs host a robust in-plane topological polarization network exhibiting a meron–antimeron texture. This network remains stable and retains its topological characteristics even under an applied out-of-plane electric field. Therefore, the complex ferroelectric response in twisted h-BN is governed predominantly by this intrinsic electromechanical coupling mechanism rather than by random defect-induced pinning effects.

Regardless of the specific stabilization mechanism, both studies reveal that the morphology and characteristic length scales of polar domains (AA and AB/BA regions) in twisted bilayer h-BN can be continuously modulated by an external electric field [85, 145], offering practical strategies for manipulating polar topologies. Similar behaviour is observed in twisted bilayer $CuInP_2S_6$ (CIPS), where both electric fields and strain can effectively tune the domain morphology and size within the moiré pattern. The domain size, thermal stability, and polarization lifetime exhibit strong dependence on twist angle and temperature. This sensitivity arises from stacking-dependent switching energy barriers and the corresponding changes in switching kinetics under thermal fluctuations [89]. Beyond twisted h-BN and CIPS, manipulation of topological polar structures is widely reported in nonplanar geometries and ferroelectric heterostructures. For instance, PbX (X = S, Se, Te) monolayers exhibit reversible ferroelectric phase transitions under tensile or shear strain. MLMD with finite-element simulations show that inhomogeneous strain fields generated by external mechanical loads, such as tip indentation or bubble inflation, can induce a variety of skyrmion-like polarization textures on the resulting topologically deformed surfaces, including vortices, antivortices, and flux-closure patterns [150]. In $(PbTiO_3)_{10}/(SrTiO_3)_{10}$ ferroelectric superlattices, three-dimensional topological polar structures are strongly coupled to thermal fluctuations and compositional doping. Upon heating, the polar vortex cores undergo a collective displacement, giving rise to an unusual sequence of in-plane polarization states: ferroelectric-like (FE) → antiferroelectric-like (AFE) → paraelectric (PE). Furthermore, introducing Pb into the STO layers (forming PSTO) can weaken the depolarization field in the PTO layers and thereby induce the formation of skyrmion bubbles within the 3D domain architecture [146] (**Figure 5f**).

Overall, MLMD is particularly powerful and shows superior advantages compared to traditional methods whenever polar topology is controlled by a competition among multiple factors including stacking patterns, twist angles, defects, strain and depolarization fields. Beyond serving as a scalable surrogate for first-principles calculations, MLMD functions as a controllable platform to scan this high-dimensional design space and to test how specific perturbations reshape the landscape of polar topological states.

### 3.4. Polar-Mechanical Coupling

Morphology and spatial distribution of ferroelectric domains and topological structures are sensitive to the thermal fluctuation, electric field and strain deformation, and balanced by the competition among depolarizing fields, elastic energy, and interfacial/interlayer couplings. This implies that mechanical deformation such as stress, strain and strain gradients is one of the most important and fundamental knobs for tuning ferroelectric properties. Specifically, when a ferroelectric material experiences mechanical deformation, the local polarization can rotate and reorient, to trigger domain-wall motion, domain bending, and even the nucleation and growth of new walls [132-133, 141]. Consequently, strain-polarization coupling enables mechanically controllable read/write pathways in ferroelectric devices, converting mechanical stimuli into measurable electrical signals for sensing, energy harvesting, and electromechanically tuned memory [151-152]. In low-dimensional ferroelectrics, the low out-of-plane bending stiffness makes them susceptible to pronounced deformation on flexible substrates, while the resulting strain gradients can activate flexoelectricity to drive local polarization and modulate domain nucleation and DW barriers in the absence of an external electric field [24, 153]. These effects provide an engineerable route toward low-power, bendable ferroelectric devices. Over the past decades, coupling between mechanical strain and polarization has been established across a wide range of ferroelectric materials. For instance, in 2D $D_{3d}$ crystals (e.g., monolayer $SnS_2$, silicene, phosphorene, $RhI_3$, and bilayer h-BN), theoretical studies predict and quantify in-plane flexoelectric responses [154]; the vdW ferroelectric CIPS exhibits pronounced flexo/piezoelectric coupling [24, 155]; layered α-$In_2Se_3$ nanoflakes display out-of-plane piezoelectricity [156]; and in $(PbTiO_3)_n/(SrTiO_3)_n$ heterostructures, room-temperature polar vortices and skyrmions arise inside ferroelectric bubble domains that are induced by lattice-mismatch strain [157-158].

Piezoelectric and flexoelectric coefficients are key metrics for quantifying polar–mechanical coupling, characterizing the polarization response to uniform strain and to strain gradients, respectively. Benefiting from the rapid advances in MLMD, these responses can now be evaluated efficiently and with high accuracy at larger length and time scales. *Xiang et al.* evaluated the shear flexoelectric coefficient of anatase $TiO_2$ using an artificial neural network (ANN) potential [159], obtaining a value of ~ **4 nC·m$^{-1}$** that is consistent with experimental measurements. Moreover, the flexoelectric-response curve predicted by the ANN agrees closely with DFT results, indicating that MLFFs can provide reliable predictions of flexoelectric responses [160]. Going beyond accuracy validation, *Javvaji et al.* developed a framework that combines a moment-tensor potential (MTP) [161] with a charge–dipole model and demonstrated in diamane monolayers that Janus diamane exhibits a linear piezoelectric response under tensile strain (absent in the symmetric counterpart) with an out-of-plane piezoelectric coefficient up to ~15× that of conventional Janus TMD monolayers [162]. Subsequent studies further show that, under bending deformation in Janus diamane and more complex 2D vdWs bilayers, bending-induced structural asymmetry enhances local electric fields and effective dipole moments through charge redistribution and interlayer coupling, thereby strengthening the linear polarization response to strain gradients and inducing the ultrahigh flexoelectric coefficient [163]. Notably, strain deformation need not originate from externally applied loads; it can also arise from interfacial lattice mismatch when distinct 2D crystals are vertically stacked. In vdWs heterostructures and bilayer TMDs, such interfacial strain couples strongly with polarization behavior, driving cooperative atomic rearrangements and spontaneous curvature, as observed in $WSe_2$, graphene/BN and $MoS_2/MoSe_2$ systems. The resulting curvature fields can propagate across interfaces and are poised to influence electronic structure, magnetism, monolayer ferroelectricity and optoelectronic responses [164-165]. Collectively, these benchmarks show that MLMD can not only predict piezoelectric and flexoelectric coefficients in close agreement with DFT and experiment at device-relevant length and time scales, but also resolve finite-temperature structural reconstructions (e.g. local symmetry breaking, concentrated strain gradients and curvature fields) that underlie polar–mechanical coupling, thereby providing an atomistic bridge between symmetry, microstructure and effective piezo/flexo responses.

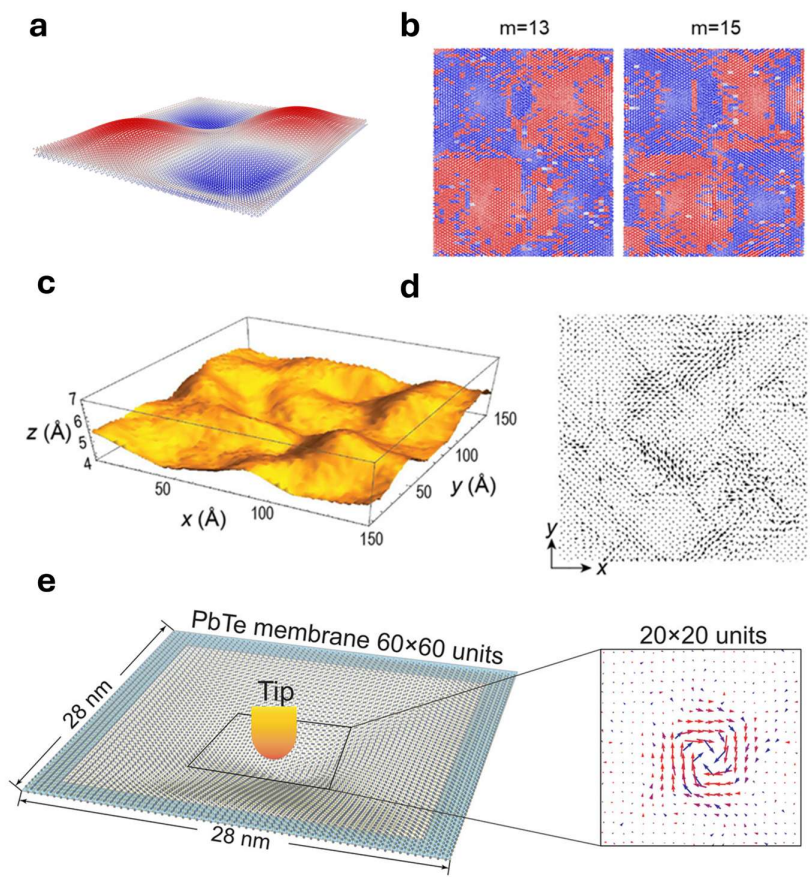

**Figure 6. Mechanical-deformation induced local polar domains and topological polar structures in two-dimensional ferroelectrics.** (a) Bubble structure in monolayer $\alpha$-In$_2$Se$_3$. (b) Polarization distributions of the bubble model in monolayer $\alpha$-In$_2$Se$_3$ at different amplitudes ($m$ = 13, 15). Reprinted by permission [25]. Copyright 2023, American Chemical Society. (c) Average surface morphology of a free-standing monolayer GeSe film at 280 K and 2.0 ps. (d) Average local ferroelectric order in monolayer GeSe at 280 K over 2.0 ps. Reprinted by permission [109]. Copyright 2021 Wiley‐VCH GmbH (e) Topological polar defects in monolayer PbTe by mechanical indentation. Reprinted by permission [150]. Copyright 2024 Wiley‐VCH GmbH.

Electromechanical responses arise not only from intrinsic lattice effects at the atomic scale, but in some systems can also be dominated by the motion, rotation, and rearrangement of DWs under strain or strain gradients. A representative example is macroscopically nonpolar SrTiO$_3$, in which ferroelastic twin walls possess weak local polarization. MLMD simulations have shown that even an ultralow strain gradient can drive the rotation and reorganization of these twin walls, thereby inducing a stable macroscopic net polarization and confirming the existence of flexo-ferroelectricity mediated by DW dynamics [166]. Similarly, in 2D sliding ferroelectrics, mechanical deformation can markedly alter polarization behavior by modulating interlayer stacking and DW evolution. For example, in bilayer h-BN and 3R-MoS$_2$, bending-induced local kinks can modify the interlayer atomic stacking and generate topological DW; accompanied by interlayer relative sliding and DW propagation, the system can undergo out-of-plane polarization reversal, exhibiting electromechanical characteristics analogous to a flexoelectric response [167]. Beyond flexoelectric responses, mechanically induced DW evolution can also markedly enhance piezoelectricity. For example, under extreme tensile strain, freestanding PbTiO$_3$ films can develop domain structures with flexible dipoles and highly mobile DWs. In this regime, the local dipoles near the DWs do not undergo abrupt switching under an external field or applied stress. Instead, they exhibit collective, cooperative small-angle rotations, thereby giving rise to an enhanced piezoelectric response. Upon further stretching, spontaneous oscillations of the 90° DWs can increase $d_{33}$ to about 250 pC/N. Moreover, the system can evolve into a polar topological structure known as a "dipole spiral". Owing to its rotational zero-energy mode, its intrinsic piezoelectric response can exceed 320 pC/N [168].

By exploiting the influence of strain gradients on polarization switching dynamics, complex strain environments can further affect the local polarization stability and phase behavior of ferroelectric materials through the control of DW nucleation and propagation. Due to their pronounced out-of-plane flexibility, two-dimensional materials readily develop ripples and bubbles under external stimuli, creating complex local strain and curvature that strongly affect polarization-switching pathways and domain evolution. For example, in monolayer α-In$_2$Se$_3$, DFT and MLMD simulations have shown that compressive strain can markedly reduce the polarization-switching barrier, whereas tensile strain significantly increases it; notably, a compressive strain of 3% can reduce the barrier to zero. Under bending, this facilitates in-plane displacements of Se atoms and thereby triggers polarization reversal. In the complex strain fields associated with ripples and bubbles, the stability and size of local polar domains can be further tuned by controlling the local curvature and temperature [25]. (**shown in Figure 6a,b**). Similarly, in ferroelastic–ferroelectric monolayer GeSe, out-of-plane ripples act as dynamic sources of random stress and profoundly alter both the phase-transition behavior and domain-switching kinetics of the material. Because ripples can stabilize short-range ferroic order in the high-temperature phase, while endowing it with stronger ferroicity and a longer lifetime, these preformed locally ordered regions can more readily evolve and expand into long-range order during cooling, thereby increasing the macroscopic phase-transition temperature. Meanwhile, under external loading, the spatially and temporally nonuniform local stress fields introduced by ripples disrupt the originally highly correlated cooperative domain switching in the ideal two-dimensional lattice, transforming the switching process into a ripple-driven localized evolution, in which high-strain regions preferentially participate in domain nucleation and propagation [109, 169]. (**Figure 6c,d**).

The strength of MLMD lies in its ability to uncover complex electromechanical coupling mechanisms in known structures while also enabling the inverse design of functional materials. For example, *Wang et al.* employed MLFFs to perform high-throughput molecular dynamics simulations on lateral h-BN/graphene heterostructures and showed that their mechanical, piezoelectric, and dielectric responses can be broadly tuned by varying the concentration and spatial arrangement of graphene domains, making the piezopotential highly designable [170]. Also as discussed in Section 3.3, *Xu et al.* demonstrate that, in 2D PbX thin films, nonuniform strain fields generated by complex mechanical loading can trigger reversible strain-induced ferroelectric phase transitions and enable the designed formation of various skyrmion-like polar topological structures on 2D surfaces, including vortices, antivortices, and flux-closure patterns [150] (**Figure 6e**).

Together, these studies highlight that MLMD serves not merely as a tool for explaining the microscopic origin of complex electromechanical phenomena, but as a practical framework for integrating local strain engineering with topological-structure control or high-throughput screening toward the rational design of novel electromechanical functional materials.

## 4. Challenge and Future Outlook

Although a promising frontier is emerging at the intersection of machine learning force field and ferroelectric physics, several associated issues remain to be addressed. We highlight key challenges and opportunities below and suggest areas where machine learning can be further leveraged.

### 4.1. Long-range Interaction

A primary and arguably most fundamental challenge in modelling ferroelectrics is the accurate treatment of long-range Coulomb interactions. In these materials, long-range electrostatics is not just one contribution among many; it often drives spontaneous polarization, shapes depolarization fields and domain structures, and controls the response of surfaces and interfaces. Within the soft-mode framework, a transverse optical phonon mode softens as the Curie temperature is approached, reflecting a lattice-dynamical instability arising from the delicate interplay between short-range repulsion and long-range Coulomb forces. Consequently, electrostatic effects extend far beyond a single unit cell and are essential for describing surface bound charges, depolarization fields, and polar discontinuities. For instance, a ferroelectric single crystal generates bound charges at its surfaces, producing a depolarization field that opposes the bulk polarization—an explicitly long-range effect.

By contrast, most current MLFFs are trained on quantum-mechanical reference data in relatively small supercells (typically < 100 atoms) and adopt a local-cutoff approximation, in which the atomic energy contribution is determined solely by a finite-radius neighborhood. In practice, cutoff radii are often on the order of 4–6 Å, which severely constrains the explicit range of electrostatic interactions encoded in the model. The impressive successes of MLFFs in ferroelectric-related problems to date largely concern bulk or 2D systems and phenomena where long-range electrostatics is screened or not dominant. The limitations of a purely local description become acute in systems with charged domain walls, polar-discontinuous heterostructures, nanopolar regions, or order–disorder ferroelectric transitions, where depolarization fields and electrostatic boundary conditions play a central role.

Addressing long-range interactions is therefore a core focus of current MLFF research. Promising strategies can be grouped into three broad paradigms. The first is physically motivated hybrid models, which explicitly decompose the total energy into long-range and short-range parts. Long-range terms (e.g. electrostatics, multipole–multipole interactions) are treated with analytic or semi-empirical models, while an ML potential captures the remaining short-range, many-body contributions. For example, the Polarizable Atom Interaction Neural Network (PAINN) predicts atomic charges and dipoles (or higher multipoles), which then interact through the self-consistent electrostatic field, thereby capturing many-body polarization effects [81]. The second paradigm is architectural solutions, in which the ML model itself is modified to propagate information over larger length scales (for instance via message passing, attention, or hierarchical pooling), so that long-range correlations emerge without an explicit analytic electrostatic term [171]. The third is specialized long-range descriptors, which design atomic environment representations with an intrinsically extended spatial reach, replacing standard short-range descriptors [172].

Each paradigm faces trade-offs among accuracy, transferability and computational efficiency, and none has yet become a standard choice for ferroelectric MLFFs. In practice, robust modeling of charged interfaces, depolarization fields and polar discontinuities remains challenging. A likely direction is the development of hybrid schemes that combine physically motivated electrostatic terms with long-range-capable architectures (e.g. attention or equivariant message passing) to account for residual quantum-mechanical long-range effects. The ultimate goal is a model class that is both computationally efficient and physically faithful, able to treat phenomena ranging from depolarization fields and polar discontinuities to nanoscale domain formation within a single MLFF framework.

## 4.2. Multiferroic Materials and Magnetoelectric Coupling

A second frontier concerns multiferroics and the accurate description of magnetoelectric coupling. Multiferroic materials exhibit two or more primary ferroic orders − typically ferroelectricity, ferromagnetism, and/or ferroelasticity−with the most technologically appealing combination being the coexistence of ferroelectricity and ferromagnetism. Their defining functionality is magnetoelectric coupling, namely the ability to control polarization with a magnetic field or magnetization with an electric field. Conventional MLFFs, such as Behler–Parrinello neural networks and Gaussian approximation potentials, treat the total energy as a function of atomic coordinates in non-magnetic systems. For multiferroics (e.g. $BiFeO_3$ and $BiMnO_3$), however, the total energy $E_{total}(R_i, \{S_i\})$ depends on both atomic positions $R_i$ and electronic spin configurations $\{S_i\}$. Conceptually, it can be decomposed as

$$E_{total}(R_i, \{S_i\}) = E_{lattice}(R_i) + E_{spin}(\{S_i\}) + E_{spin\text{-}lattice}(R_i, \{S_i\})$$

where the critical term is the spin–lattice coupling $E_{spin\text{-}lattice}$, which couples structural distortions and phonon modes to magnetic exchange, anisotropy and Dzyaloshinskii–Moriya interactions. Recent work has proposed several strategies to incorporate spin degrees of freedom into MLFF frameworks:

- **Spin-constrained (spin-averaged) potentials.**

  This is the most straightforward extension, mainly suited to collinear magnetic states (e.g. uniform ferromagnets or simple antiferromagnets). The spin configuration is treated as a fixed, global property, and a separate MLFF is trained for each chosen magnetic order. This approach is easy to implement within existing MLFF frameworks, but suffers from poor transferability: a potential trained on one magnetic state cannot describe another, and non-collinear textures (e.g. skyrmions) or magnetic phase transitions are inaccessible.

- **Spin-dependent MLFFs.**

  Here one seeks a unified model that explicitly depends on spin configurations, learning the energy as $E = E(R_i, \{S_i\})$. Atomic spin vectors are provided as inputs alongside atomic coordinates and species, with network architectures that process geometric and spin information in separate channels and couple them in deeper layers (e.g. the DeepSPIN framework with spin-dependent embeddings [173-174]). Such models in principle enable molecular dynamics or Monte Carlo simulations in which magnetic order evolves with temperature, and energy differences between spin-spiral states can be used to extract effective Heisenberg and Dzyaloshinskii-Moriya parameters. However, the data requirements are severe: training demands extensive DFT datasets spanning a wide variety of spin configurations and structures, which is computationally expensive, especially for configurations far from equilibrium.

- **Learned spin Hamiltonians**

  A more abstract approach focuses directly on magnetic interactions rather than the total energy. Here, a deep neural network learns the parameters of an effective classical spin Hamiltonian (e.g. Heisenberg + Dzyaloshinskii–Moriya terms) from DFT data. Crucially, exchange parameters $J_{ij}$ and Dzyaloshinskii–Moriya vectors $D_{ij}$ are modeled as functions of the local chemical and structural environment around each pair $i$–$j$. This yields direct physical insight into how structure and chemistry shape magnetic interactions. The price is that the lattice is usually treated as rigid, dynamic spin–lattice coupling is neglected, and the approach relies on the assumed functional form of the spin Hamiltonian, which may miss multi-spin interactions or itinerant magnetism.

Despite rapid progress, several challenges remain before spin-aware MLFFs can be broadly deployed in multiferroic and magnetoelectric systems. The most critical bottleneck is data scarcity: the combined configurational space of atomic positions and spin degrees of freedom is enormous, rendering systematic DFT-level sampling prohibitively expensive. Constructing representative training datasets that capture both equilibrium and far from equilibrium spin–lattice states is still largely manual. From a simulation perspective, developing efficient algorithms that concurrently evolve atomic positions and spins (e.g. coupled spin–lattice dynamics at finite temperature and under external fields) within an ML framework remains an additional hurdle. Progress in these directions will be pivotal for using MLFFs as quantitative tools for the rational design of next generation multiferroic, spintronic and magnetoelectric devices.

## 4.3. Pre-training Large Atomistic Model

A third point is large atomistic model (LAM). High-entropy materials stabilize single-phase structures despite containing multiple elements in near-equimolar proportions, leading to enhanced mechanical, thermal, electrochemical, and functional properties. Recently, some researches show that the high-entropy strategy has emerged as an effective and flexible approach for boosting ferroelectric and dielectric properties in high-entropy ferroelectrics via the delicate design of local polarization configurations and other intrinsic effects [175-176]. The combinatorial complexity of vast chemical and configurational spaces in high-entropy ferroelectrics, however, makes exhaustive exploration impractical. Traditional MLFFs are specialized models that trained on extensive DFT data generated for a specific chemical high-entropy system, therefore they have two major issues: (i) Each new composition requires generating a new large DFT dataset and training a new MLFF from scratch-prohibitively expensive. For example, An MLFF trained on $(Ca_{0.2}Sr_{0.2}Ba_{0.2}Pb_{0.2}Nd_{0.1}Na_{0.1})Bi_4Ti_4O_{15}$ fails for $Pb(Ni_{0.185}In_{0.074}Zr_{0.037}Ti_{0.259}Nb_{0.444})O_3$ or a subsystem; (ii) Conventional fixed-length descriptors in MLFF struggle to uniquely represent the extreme diversity of local environments in high-entropy ferroelectrics (atoms with five more different neighbor types), leading to "degeneracy" where different environments map to similar descriptors. Pre-trained LAMs represent a transformative paradigm at the intersection of artificial intelligence and materials science. LAMs have seen nearly all elements and many combinations during pre-training on vast databases like the Materials Project. They can make reasonable predictions for unseen elemental mixtures by leveraging learned elemental embeddings and interaction patterns. This enables rapid screening across the vast high-entropy space with minimal new data. In recent years, there has been significant development in LAMs with broad coverage, such as DPA-3 [177], MatterSim [178], ORB [179], MACE [180], CHGNet [181], and M3GNet [182], among others. With their sophisticated model architecture, universal force fields exhibit exceptional generalizability after training using millions of first-principles calculations covering a wide range of materials. Ideally, universal force fields are applicable to high-entropy ferroelectrics without the need for additional training and have seemingly addressed the above issues of the traditional MLFFs.

However, despite their high generalizability, universal force fields still face several key challenges. First, their energy error typically ranges from several tens to hundreds of meV/atom, which may be inadequate for simulating disordered electric dipole in high-entropy ferroelectrics where first-principles accuracy-often requiring errors within a few meV/atom-is essential. Additionally, the complex neural network architectures (e.g., transformers, deep graph networks) required for their generality incur substantially higher computational cost per atom compared to MLFFs. This limits their practical use for large-scale MD simulations involving hundreds of thousands of atoms or long-time scales. Due to these limitations, universal LAMs remain unsuitable for accurate large-scale simulations of high-entropy systems. Therefore, instead of pursuing a single perfect model, a more practical approach may be to use prior knowledge to derive fast, system-specific force fields from a universal force field, tailored for high-entropy applications.

# 5. Conclusion

In summary, these developments indicate that MLMD has evolved from a computationally efficient surrogate of DFT at extended length and time scales to a versatile framework capable of quantitatively addressing polarization switching, domain-wall motion, topological polar textures, and curvature-driven polar-mechanical coupling in ferroics. At the same time, robust treatment of long-range electrostatics and spin-lattice coupling, together with pre-trained large atomistic models to chemically and configurationally complex ferroic systems, remain key bottlenecks. We anticipate that progress in hybrid long-range MLFFs, spin-aware force fields, and pre-training large atomistic model explicitly designed to target ferroic order parameters and device-relevant performance metrics will be crucial. Addressing these challenges will position MLMD-based force fields as a predictive and quantitative tool for the rational design and optimization of ferroelectric and multiferroic devices.

# Author Contributions

L.K. proposed the idea and writing framework of this Perspective. D.B. drafted Chapter 1 (Introduction), Chapter 2 (MLMD: From Concept to Implementation), and Chapter 3 (Ferroelectric Research Progress Enabled by MLMD). R.H. drafted Chapter 4 (Challenges and Future Outlook). J.L. revised the manuscript for language polishing and grammatical corrections. All authors reviewed and approved the final version of the manuscript. D.B. and R.H. contributed equally to this work and are recognized as co-first authors.

# Acknowledgement

D.B. acknowledges support from the Queensland University of Technology Postgraduate Research Award (QUTPRA) at QUT. L.K. gratefully acknowledges financial support from the ARC Discovery Project (DP230101904 and DP240103085). R.H. gratefully acknowledges the Natural Science Foundation of China (12574108), the Zhejiang Provincial Natural Science Foundation of China (DG26A040001), and the Talent Hub for "AI+ New Materials" Basic Research.